\documentclass[prd,tightenlines]{revtex4}
\usepackage{epsfig}
\usepackage{graphics}
\usepackage{latexsym}
\usepackage{amsmath}
\usepackage{amssymb}
\usepackage{rotating}
\usepackage{subfigure}

\newcommand{\be}{\begin{equation}}
\newcommand{\ee}{\end{equation}}
\newcommand{\ba}{\begin{eqnarray}}
\newcommand{\ea}{\end{eqnarray}}
\newcommand{\non}{\nonumber}

\begin{document}

\title{Study on $0^+$ states with open charm in unitarized heavy
meson chiral approach}

\author{P. Wang}
\author{X. G. Wang}
\affiliation{Institute of High Energy Physics, CAS, P. O. Box
918(4), Beijing 100049, China} \affiliation{Theoretical Physics
Center for Science Facilities, CAS, Beijing 100049, China}

\begin{abstract}
We calculate the scattering amplitudes of Goldstone bosons off the
pseudoscalar D-mesons in unitarized heavy meson chiral approach. The
low energy constants appearing in $\mathcal{O}(p^2)$ chiral
Lagrangian are determined by fitting lattice simulations on $S$-wave
scattering lengths. $D_{s0}^{*}(2317)$ is obtained as a bound state
in $(S,I)=(1,0)$ $DK$ channel. Possible bound states or resonance
states in other channels are investigated as well. The quark mass
dependence of the mass and binding energy of $D_{s0}^{*}(2317)$ is
also investigated, which indicates predominately $DK$ molecular
nature.
\end{abstract}


\maketitle


\smallskip

\section{Introduction}
In the last decade, the discovery of many narrow resonances with
open charm open a new chapter in hadronic spectroscopy. Especially,
the $D_{s0}^{*}(2317)$ discovered by the BaBar
Collaboration~\cite{BaBar:2003} and $D_{s1}(2460)$ by the CLEO
Collaboration~\cite{CLEO:03} have inspired heated discussions both
experimentally and theoretically. Moreover, the Belle Collaboration
recently reported a broad $0^+$ charmed meson with mass and width
being $m_{D^{*0}_0}=2308\pm60\mathrm{MeV}$ and
$\Gamma_{D^{*0}_{0}}=276\pm99\mathrm{MeV}$,
respectively~\cite{Belle:04}. Meanwhile, the FOCUS Collaboration
reported a broad $0^+$ charmed meson with mass and width being
$m_{D^{*0}_0}=2407\pm56\mathrm{MeV}$ and
$\Gamma_{D^{*0}_{0}}=240\pm114\mathrm{MeV}$,
respectively~\cite{FOCUS:04}. Although consistent with each other
within the errors, it is still in dispute whether they are the same
particle~\cite{Bracco:05}.

Possible interpretations of $D_{s0}^{*}(2317)$ include normal
$c\bar{s}$ state~\cite{Bardeen:03}, four-quark
state~\cite{Cheng:03}, hadron molecular state~\cite{Barnes:03}, etc.
To distinguish composite from elementary particles, different
methods were proposed such as pole counting~\cite{morgan92},
scattering length and effective range~\cite{Weinberg:63}. As
emphasized in a series of paper~\cite{Hanhart:08,Cleven:11}, quark
mass dependence of a state can also provide important information on
its nature.

Effective field theories(EFTs) have been proven very successful in
studying low energy hadron physics~\cite{Weinberg79}. In light meson
sector, chiral perturbation theory is an expansion in powers of
external momenta and masses of Goldstone
bosons~\cite{Gasser84,Gasser85}. In high energy region or for large
quark masses, chiral amplitudes violate unitarity severely. In
addition, chiral expansion up to a given finite order does not
contain resonance or bound state, which may modify the results of
physical variables from perturbation theory significantly.
Therefore, unitarized model was introduced to high energy region, in
which lower lying scalar and vector resonances can be dynamically
generated. Following the same spirit, heavy meson chiral
perturbation theory (HMChPT) was
proposed~\cite{Burdman:1992gh,Wise:1992hn,Yan:1992gz}, and
unitarization method was applied to some phenomenological
analysis~\cite{Lutz:08,Hofmann:04,Guo:06,Guo:07,Guo:09}.

Lattice gauge theory is another powerful tool to study strong
interactions. Lattice simulations are usually performed at
unphysical quark masses, or equivalently at larger pion masses.
Recently, lattice results for the charmed meson-light hadron
scattering lengths are given at several chosen values of
$M_{\pi}/F_{\pi}$~\cite{Liu:2008rz}. These progresses can be used to
make up the lack of experimental data on scattering processes. These
lattice data can be used to determine the low energy constants in
perturbative scattering amplitudes~\cite{Liu:09}.

$D_{s0}^{*}(2317)$ as well as other possible charmed particles were
investigated with the unitarized heavy meson chiral approach by
studying the scattering lengths of charmed mesons and Goldstone
bosons in Ref.~\cite{Cleven:11,Guo:09}. The quark mass dependence of
the poles has an interesting behavior and provides a good way to
understand the structure of the obtained poles. In their
calculation, the large $N_C$ approximation and the mass and width of
$D_{s0}^{*}(2317)$ as input are used to determine the low energy
constants. In this paper, we use the similar approach to
reinvestigate the charmed mesons scattering off light mesons. The
difference between our treatment and theirs is that we do not apply
the large $N_C$ approximation. This is because in the real world,
$N_C$ is 3. Moreover, in the large $N_C$ limit, the mass and width
of the particle can be quite different from the real particle
\cite{Pelaez04,Guo11,Zheng:12}. Therefore, three additional low
energy constants (LECs) $h_0$, $h_2$ and $h_4$ appear in our case.
The parameter $h_1$ can be determined by the mass difference between
$D$ mesons. The parameter $h_0$ is obtained by the quark mass
dependence of $D$ and $Ds$ mesons from the lattice data
\cite{HPQCD:2008}. The other constants including $h_3$ and $h_5$ are
determined by fitting the lattice data of the scattering lengths of
$D$ and light mesons \cite{Liu:2008rz}. To confirm the existence of
$D_{s0}^{*}(2317)$, its mass and width are not used as input. All
the states including $D_{s0}^{*}(2317)$ will be obtained from pole
analysis on the scattering amplitudes. As a comparison, the quark
mass dependence of $D_{s0}^{*}(2317)$ is also discussed.

The paper is organized as follows. In sect. II, the effective chiral Lagrangian
up to next-to-leading order is briefly introduced. We
calculate the unitary scattering amplitudes and determine the low energy constants
by fitting lattice simulations on $S$-wave scattering lengths in sect. III.
In sect. IV, we present the possible bound states or resonant
poles in appropriate channels, and then investigate the quark mass
dependence of $D_{s0}^{*}(2317)$. Finally, We make a brief summary
in sect. V.

\section{The effective Lagrangian}

The leading order chiral Lagrangian for describing the interaction
between the Goldstone boson and the heavy pseudoscalar meson
is~\cite{Burdman:1992gh,Wise:1992hn,Yan:1992gz}
\begin{equation}
{\cal L}^{(1)} = {\cal D}_{\mu}D{\cal
D}^{\mu}D^{\dag}-\overset{_\circ}{M}_D^2 DD^{\dag}
\end{equation}
with $D=(D^0,D^+,D_s^+)$. The covariant derivative is
\begin{eqnarray}
{\cal D}_{\mu}D^{\dag}&=&(\partial_{\mu}+\Gamma_{\mu})D^{\dag}\ ,\non\\
\Gamma_{\mu}&=&\frac{1}{2}(u^{\dag}\partial_{\mu}u+u\partial_{\mu}u^{\dag})\ ,
\end{eqnarray}
where
\begin{equation}
U=\exp(\frac{\sqrt{2}i\phi}{F})\ ,\ \ \ \ u^2=U\ ,
\end{equation}
with $\phi$ containing the Goldstone boson fields,
\begin{equation}
\phi(x)=\left(
          \begin{array}{ccc}
            \frac{1}{\sqrt{2}}\pi^0+\frac{1}{\sqrt{6}}\eta & \pi^+ & K^+ \\
            \pi^- & -\frac{1}{\sqrt{2}}\pi^0+\frac{1}{\sqrt{6}}\eta & K^0 \\
            K^- & \bar{K}^0 & -\frac{2}{\sqrt{6}}\eta \\
          \end{array}
        \right)\ .
\end{equation}
$F$ is the Goldstone boson decay constant in the chiral limit, which
we will identify with the pion decay constant, $F=92.4\mathrm{MeV}$.

The strong interaction part of NLO chiral Lagrangian reads
\begin{eqnarray}\label{eq:L2str}
{\cal L}^{(2)}_{\rm str.} &\!\!=&\!\! D \bigl(
-h_0\langle\chi_+\rangle - h_1\chi_+ + h_2\left\langle
u_{\mu}u^{\mu} \right\rangle - h_3u_{\mu}u^{\mu}
\bigr) {\bar D} \non\\
& + & {\cal D}_{\mu}D \bigl( h_4\langle u^{\mu}u^{\nu}\rangle - h_5
\{u^{\mu},u^{\nu}\} - h_6 [u^{\mu},u^{\nu}] \bigr) {\cal
D}_{\nu}{\bar D} ,
\end{eqnarray}
where $<>$ stands for the trace of the $3\times 3$ matrices, and
\begin{eqnarray}
\chi_+&=&u^{\dag}\chi u^{\dag}+u\chi u\ ,\non\\
u_{\mu}&=&iu^{\dag}{\cal D}_{\mu}U u^{\dag}\ .
\end{eqnarray}
with
\begin{equation}
\chi = 2B\cdot {\rm diag}\left(m_u,m_d,m_s\right)\ .
\end{equation}
The $h_1$ term is a little different from ~\cite{Guo:09} in order
that the term $D\langle\chi_+\rangle\bar{D}$ will completely
disappear in large $N_C$ limit~\cite{Lutz:08}. The corresponding
coefficients $C_1$ are also modified (see Tab.~\ref{tab:Vstu}). The
term proportional to $h_0$ leads to a singlet contribution to the
$D$-meson masses which depends linearly on the light quark masses,
and is the heavy meson analog of the pion-nucleon sigma
term~\cite{Jenkins:94}. The $h_1$ term will contribute to the
$SU(3)_V$-violating mass splitting amongst $D$ mesons. The masses of
$D$ and $D_s$ mesons can be expressed as
\begin{eqnarray}\label{mass}
M_D^2&=&\overset{_\circ}{M}_D^2+4h_0B(m_u+m_d+m_s)+4h_1B\hat{m}\ ,\non\\
M_{D_s}^2&=&\overset{_\circ}{M}_D^2+4h_0B(m_u+m_d+m_s)+4h_1B m_s\ ,
\end{eqnarray}
from which we can determine
\begin{equation}
h_1=\frac{M_{D_s}^2-M_D^2}{4B(m_s-\hat{m})}=\frac{M_{D_s}^2-M_D^2}{4(M_K^2-M_{\pi}^2)}=0.427\ ,
\end{equation}
where $\hat{m}=(m_u+m_d)/2$ and the mass relations of Goldstone bosons from leading order chiral expansion,
\begin{equation}\label{mass-relation}
M_{\pi}^2=2B\hat{m}\ ,\ \ \ M_K^2=B(\hat{m}+m_s)\ ,\ \ \ M_{\eta}^2=\frac{2}{3}B(\hat{m}+2m_s)\ ,
\end{equation}
are used. We can simply estimate the value of $h_0$ to be 0.055 according to the slope of the extrapolation curve from lattice~\cite{HPQCD:2008}.

\begin{table*}[t]
\begin{center}
\renewcommand{\arraystretch}{1.3}
\begin{tabular}{|ll|ccccc|}\hline\hline
$(S,I)$         &          Channel            &         $C_{LO}$      &            $C_1$                      &    $C_{35}$           &   $C_0$      & $C_{24}$ \\ \hline%
$(-1,0)$        & $D{\bar K}\to D{\bar K}$    &         $-1$          &          $3M_K^2$                     &     $-1$              & $-M_{K}^2$   &  $-1$    \\%
$(-1,1)$        & $D{\bar K}\to D{\bar K}$    &          1            &         $-3M_K^2$                     &       1               & $-M_{K}^2$   &  $-1$    \\%
$(0,{\frac12})$ & $D\pi\to D\pi$              &         $-2$          &       $-3M_\pi^2$                     &       1               & $-M_{\pi}^2$ &  $-1$    \\%
                & $D\eta\to D\eta$            &          0            &        $-M_\pi^2$                     &  ${\frac13}$          & $-M_{\eta}^2$&  $-1$    \\%
                & $D_s{\bar K}\to D_s{\bar K}$&         $-1$          &         $-3M_K^2$                     &       1               & $-M_{K}^2$   &  $-1$    \\%
                & $D\eta\to D\pi$             &          0            &       $-3M_\pi^2$                     &       1               &     0        &    0     \\%
                & $D_s{\bar K}\to D\pi$       & $-\frac{\sqrt{6}}{2}$ & $-\frac{3\sqrt{6}}{4}(M_K^2+M_\pi^2)$ & $\frac{\sqrt{6}}{2}$  &     0        &    0     \\%
                & $D_s{\bar K}\to D\eta$      & $-\frac{\sqrt{6}}{2}$ & $\frac{\sqrt{6}}{4}(5M_K^2-3M_\pi^2)$ & $-\frac{\sqrt{6}}{6}$ &     0        &    0     \\%
$(0,{\frac32})$ & $D\pi\to D\pi$              &          1            &       $-3M_\pi^2$                     &       1               & $-M_{\pi}^2$ &  $-1$    \\%
$(1,0)$         & $DK\to DK$                  &         $-2$          &        $-6M_K^2$                      &       2               & $-M_{K}^2$   &  $-1$    \\%
                & $D_s\eta\to D_s\eta$        &          0            &     $-2(2M_K^2-M_\pi^2)$              & ${\frac43}$           & $-M_{\eta}^2$&  $-1$    \\%
                & $D_s\eta\to DK$             &     $-\sqrt{3}$       & $\frac{\sqrt{3}}{2}(3M_\pi^2-5M_K^2)$ & $\frac{\sqrt{3}}{3}$  &     0        &    0     \\%
$(1,1)$         & $D_s\pi\to D_s\pi$          &          0            &            0                          &       0               & $-M_{\pi}^2$ &  $-1$    \\%
                & $DK\to DK$                  &          0            &            0                          &       0               & $-M_{K}^2$   &  $-1$    \\%
                & $DK\to D_s\pi$              &          1            &    $-{\frac32}(M_K^2+M_\pi^2)$        &       1               &     0        &    0     \\%
$(2,\frac12)$   & $D_sK\to D_sK$              &          1            &          $-3M_K^2$                    &       1               & $-M_{K}^2$   &  $-1$    \\\hline\hline%
\end{tabular}
\caption{\label{tab:Vstu}{\small The coefficients in the scattering
amplitudes. Here, $S$ ($I$) denotes the total strangeness (isospin)
of the two--meson system.}}
\end{center}
\end{table*}
\section{Scattering amplitudes and unitarization}
The perturbative chiral amplitudes up to NLO can be easily obtained.
Besides the terms in Ref.~\cite{Guo:09} where large $N_C$ suppressed
ones are omitted, there are two additional terms (last two terms in
Eq.~(\ref{amp})). Although suppressed in large $N_C$ limit, the
contributions from $h_0$, $h_2$ and $h_4$ terms may not be
negligible since we are working at $N_C=3$. On the other hand,
complete large $N_C$ analysis in light meson sector shows that poles
will move far away from their physical positions as increasing
$N_C$~\cite{Pelaez04,Guo11,Zheng:12}. Therefore, in this paper we
include the complete tree level amplitude with definite strangeness
and isospin up to $\mathcal{O}(p^2)$, which can be written as
\begin{eqnarray}\label{amp}
T(s,t,u)&=&T^{(1)}(s,t,u)+T^{(2)}(s,t,u)\non\\
&=&\frac{C_{LO}}{4F^2}(s-u)+\frac{2C_1}{3F^2}h_1+\frac{2C_{35}}{F^2}H_{35}(s,t,u)\non\\
&+&\frac{4C_0}{F^2}h_0+\frac{2C_{24}}{F^2}H_{24}(s,t,u)\ ,
\end{eqnarray}
where the subscripts denote the chiral dimension and the functions $H_{35}$ and $H_{24}$ are expressed as
\begin{eqnarray}
H_{35}(s,t,u)&=&h_3 p_2\cdot p_4+h_5(p_1\cdot p_2 p_3\cdot p_4 + p_1\cdot p_4 p_2\cdot p_3)\ ,\non\\
H_{24}(s,t,u)&=&2h_2 p_2\cdot p_4+h_4(p_1\cdot p_2 p_3\cdot p_4 + p_1\cdot p_4 p_2\cdot p_3)\ .
\end{eqnarray}
Here, we adopt the same convention for the isospin decompositions as
Ref.~\cite{Guo:09}. The coefficients in all the amplitudes are given
in Tab.~\ref{tab:Vstu}. We also dropped the $h_6$ term as in
Ref.~\cite{Guo:09} since it is suppressed by one order due to the
commutator structure. The tree level amplitudes can be projected to
the $S$-wave by using
\begin{equation}\label{Swave}
V^{(S,I)}_{ij}(s)=\frac{1}{2}\int_{-1}^{1}d\cos\theta T^{(S,I)}_{ij}\big(s,t(s,\cos\theta),u(s,\cos\theta)\big)\ ,
\end{equation}
where
\begin{eqnarray}
u(s,\cos\theta)&=&m_1^2+m_4^2-\frac{1}{2s}[s+m_1^2-m_2^2][s+m_4^2-m_3^2]\non\\
&&-\frac{1}{2s}\sqrt{\lambda(s,m_1^2,m_2^2)\lambda(s,m_3^2,m_4^2)}\cos\theta\ ,
\end{eqnarray}
with
\begin{equation}
\lambda(s,m_i^2,m_j^2)=[s-(m_i+m_j)^2][s-(m_i-m_j)^2]\ .
\end{equation}

In~\cite{Oller:2000fj,Oller:2000ma,Oller:1998zr}, a general method was proposed to construct scattering amplitudes satisfying unitarity, i.e.
\begin{equation}\label{unitarization}
T(s)=V(s)[1-G(s)\cdot V(s)]^{-1}\ ,
\end{equation}
where $V$ is a matrix whose elements are given by Eq.~(\ref{Swave}) and $G$ is a diagonal matrix with the element being a two-meson integral
\begin{equation}
G_{ii}(s)=i\int\frac{d^4 q}{(2\pi)^4}\frac{1}{q^2-m_1^2+i\epsilon}\frac{1}{(p_1+p_2-q)^2-m_2^2+i\epsilon}\ ,
\end{equation}
with $m_1$ and $m_2$ the masses of the particles appearing in the loop. The analytic expression of $G_{ii}(s)$ can be expressed by~\cite{Oller:1998zr}
\begin{eqnarray}
G_{ii}&=&\frac{1}{16\pi^2}\bigl\{ a(\mu)+\log\frac{m_1^2}{\mu^2}+\frac{\Delta-s}{2s}\log\frac{m_1^2}{m_2^2}\non\\
&+&\frac{\sigma}{2s}[\log(s-\Delta+\sigma)+\log(s+\Delta+\sigma)-\log(-s+\Delta+\sigma)-\log(-s-\Delta+\sigma)]\bigr\}\ ,\non\\
\end{eqnarray}
where
\begin{equation}
\sigma=[-(s-(m_1+m_2)^2)(s-(m_1-m_2)^2)]^{1/2}\ ,\ \ \ \Delta=m_1^2-m_2^2\ .
\end{equation}
$a(\mu)$ is the subtraction constant with $\mu$ the regularization
scale. In the numerical calculation, we tried three possible values
of $a(m_D)$ estimated in Ref.~\cite{Guo:06}. Here, we did not use the the inverse
amplitude method (IAM) to get the unitary scattering amplitude. For the light meson scattering, the
divergence of one loop contributions from the leading order ($\mathcal{O}(p^2)$) Lagrangian can be subtracted from the next
leading order ($\mathcal{O}(p^4)$) tree diagram resulting in the renormalized low energy constants. Since there are a lot of
experimental and lattice data, the low energy constants for light mesons are well determined. However,
for heavy meson, if we use the IAM to get the unitary scattering amplitude, the low energy constants
at $\mathcal{O}(p^3)$ are needed to cancel the divergence. The current data for heavy meson are not enough to determine all the
constants at $\mathcal{O}(p^2)$ and $\mathcal{O}(p^3)$ very well. Therefore, in this paper, we apply the $T$-matrix formalism proposed
in Refs.~\cite{Oller:2000fj,Oller:2000ma,Oller:1998zr}, where only one parameter, the subtraction constant $a(\mu)$ was introduced.
From Eq.~(\ref{VG}), one can see that $G(s_{thr})$ is up to $\mathcal{O}(p^1)$. If we include a linear dependence term of $m_{\pi}^2$
in $a(\mu)$, as pointed out in Ref.~\cite{Cleven:11}, this higher order contribution would not change the general features of the results.

The $S$-wave scattering
length is defined as
\begin{equation}\label{a0}
a_0=-\frac{1}{8\pi(M_1+M_2)}T_{\rm thr}\ ,
\end{equation}
with $M_1$ and $M_2$ denoting the masses of the scattered heavy and
light mesons, respectively. $T_{thr}$ is the unitarized amplitude at
threshold, $s_{thr}=(M_1+M_2)^2$, which can be obtained from
Eq.(\ref{unitarization}), with
\begin{eqnarray}\label{VG}
V(s_{thr})&=&\frac{1}{F^2}\big[C_{LO}M_1 M_2+\frac{2C_1}{3}h_1+2C_{35}(h_3 M_2^2+2h_5 M_1^2 M_2^2)\nonumber\\
&&\ \ \ \ \ \ \ \ \ \ \ \ \ \ +4C_0 h_0+4C_{24}(h_2 M_2^2+h_4 M_1^2 M_2^2)\big]\ ,\nonumber\\
G(s_{thr})&=&\frac{1}{16\pi^2}\big[a(\mu)+\frac{1}{M_1+M_2}(M_1\ln\frac{M_1^2}{\mu^2}+M_2\ln\frac{M_2^2}{\mu^2})\big]\ .
\end{eqnarray}

\begin{table*}[t]
\begin{center}
\renewcommand{\arraystretch}{1.3}
\begin{tabular}{|c|c|c|c|}\hline\hline
      \          &       Fit I       &        Fit II      &    Fit III   \\ \hline
   $a(m_D)$      &     $-0.373$      &      $-0.630$      &   $-0.864$  \\ \hline
     $h_2$       & $-0.216\pm0.022$  &  $-0.195\pm0.028$  &  $-0.127\pm0.025$ \\
     $h_3$       & $0.393\pm0.180$   &  $0.510\pm0.320$   &  $-0.015\pm0.240$ \\
     $h_4$       & $0.061\pm0.007$   &  $0.056\pm0.007$   &  $0.038\pm0.005$ \\
     $h_5$       & $-0.001\pm0.020$  &  $0.032\pm0.014$   &  $0.172\pm0.060$ \\ \hline
$\chi^2_{d.o.f}$ & $43.4/12=3.6$     &  $43.6/12=3.6$     &  $42.1/12=3.5$ \\ \hline
\end{tabular}
\caption{\label{tab:LECs}{\small The fit results on LECs
corresponding to different values of $a(m_D)$ taken from
Ref.~\cite{Guo:06}. $h_2$ and $h_3$ are dimensionless, while $h_4$
and $h_5$ are in unit of $\mathrm{GeV}^{-2}$.}}
\end{center}
\end{table*}
\begin{table*}[t]
\begin{center}
\renewcommand{\arraystretch}{1.5}{\scriptsize
\begin{tabular}{|ll|rrrl|r|}\hline\hline
$(S,I)$          &               Channel        &    LO     &          NLO                 &     UChPT                   &   CUChPT        & Lattice~\cite{Liu:2008rz}~ \\ \hline%
$(-1,0)$         & $D{\bar K}\to D{\bar K}$     &   0.36    &    $0.54^{+0.14}_{-0.15}$    &   $-1.24^{+0.41}_{-1.04}$   &                 &               \\
$(-1,1)$         & $D{\bar K}\to D{\bar K}$     &  $-0.36$  &   $-0.49^{+0.15}_{-0.14}$  &   $-0.21^{+0.03}_{-0.03}$   &                 &    $-0.23(4)$ \\
$(0,{\frac12})$  & $D\pi\to D\pi$               &   0.24    &    $0.23^{+0.01}_{-0.01}$    &    $0.41^{+0.05}_{-0.04}$     &   $0.39^{+0.04}_{-0.04}$          &               \\
                 & $D\eta\to D\eta$             &      0    &   $-0.06^{+0.07}_{-0.07}$  &   $-0.05^{+0.06}_{-0.05}$   & $-1.48^{+0.26}_{-0.44}+i0.04^{+0.20}_{-0.01}$   &               \\
                 & $D_s{\bar K}\to D_s{\bar K}$ &   0.36    &    $0.23^{+0.15}_{-0.15}$    &    $0.58^{+0.78}_{-0.47}$     & $-0.67^{+0.07}_{-0.44}+i0.10^{+0.38}_{-0.02}$   &               \\
$(0,{\frac32})$  & $D\pi\to D\pi$               &  $-0.12$  &   $-0.13^{+0.01}_{-0.01}$  &   $-0.10^{+0.01}_{-0.01}$   &                 & $-0.16(4)$    \\
$(1,0)$          & $DK\to DK$                   &   0.72    &    $0.46^{+0.27}_{-0.27}$    &   $-1.99^{+1.22}_{-0.39}$   & $-0.73^{+0.18}_{-0.55}$         &               \\
                 & $D_s\eta\to D_s\eta$         &      0    &   $-0.16^{+0.23}_{-0.23}$  &   $-0.11^{+0.20}_{-0.08}$   & $-0.35^{+0.07}_{-0.09}+i0.05^{+0.11}_{-0.03}$   &               \\
$(1,1)$          & $D_s\pi\to D_s\pi$           &      0    &    $0.003^{+0.001}_{-0.002}$   &    $0.003^{+0.001}_{-0.002}$    &  $0.010^{+0.008}_{-0.002}$           & $0.00(1)$     \\
                 & $DK\to DK$                   &      0    &    $0.02^{+0.02}_{-0.01}$    &    $0.03^{+0.01}_{-0.02}$     & $-0.52^{+0.08}_{-0.02}+i0.22^{+0.31}_{-0.11}$   &               \\
$(2,{\frac12})$  & $D_sK\to D_sK$               &  $-0.36$  &   $-0.51^{+0.15}_{-0.14}$  &   $-0.22^{+0.03}_{-0.02}$   &                 & $-0.31(2)$    \\ \hline\hline%
\end{tabular}}
\caption{\label{tab:a0}{\small The S-wave scattering lengths from
calculations at LO and NLO (units are fm). The results using
unitarized amplitudes are also given in the two columns denoted by
UChPT and CUChPT, representing one--channel and coupled--channel
unitarized chiral perturbation theory, respectively. LECs are taken
from Fit III.}}
\end{center}
\end{table*}

There are no experimental data for the scattering of Goldstone
bosons off $D$-mesons available. However, the low energy constants
entering into NLO Lagrangian ${\cal L}^{(2)}_{\rm str.}$ can be
determined by fitting the recent lattice simulations on $S$-wave
scattering lengths~\cite{Liu:2008rz}. The lattice spacing is
$b=0.12\mathrm{fm}$. The $s$ quark mass is $80\mathrm{MeV}$, which
is consistent with its physical mass, and four ensembles are chosen
with $M_{\pi}=0.1842, 0.2238, 0.3113, 0.3752$ in lattice unit, or
equivalently $M_{\pi}=0.2925, 0.3554, 0.4943, 0.5958$ in unit of
$\mathrm{GeV}$ (These lattice data were misused in~\cite{Guo:09}).

From Eq.~(\ref{mass}), the pion mass dependence of $D$ mesons up to
$\mathcal{O}(M_{\pi}^2)$ can be expressed as
\begin{eqnarray}\label{MDMDs}
M_D(M_\pi)&=&M_D|_{phy}+\frac{2h_0+h_1}{M_D|_{phy}}(M_{\pi}^2-M_{\pi}^2|_{phy})\ ,\non\\
M_{Ds}(M_\pi)&=&M_{Ds}|_{phy}+\frac{2h_0}{M_{Ds}|_{phy}}(M_{\pi}^2-M_{\pi}^2|_{phy})\ ,
\end{eqnarray}
and from Eq.~(\ref{mass-relation}) we can get
\begin{equation}\label{MKMeta}
M_K(M_{\pi})=\overset{_\circ}{M}_K+\frac{M_{\pi}^2}{4\overset{_\circ}{M}_K}\
,\ \ \
M_{\eta}(M_{\pi})=\overset{_\circ}{M}_{\eta}+\frac{M_{\pi}^2}{6\overset{_\circ}{M}_{\eta}}\
,
\end{equation}
where $\overset{_\circ}{M}_K=486\mathrm{MeV}$ and
$\overset{_\circ}{M}_{\eta}=542\mathrm{MeV}$ are the masses of kaon
and $\eta$ in the chiral limit, respectively. The physical masses
for all mesons are taken from PDG~\cite{Amsler:2008zz}, i.e.,
$M_\pi |_{phy}=138$ MeV, $M_K |_{phy}=496$ MeV, $M_\eta |_{phy}=548$ MeV, $M_D |_{phy}=1867$ MeV,
and $M_{D_s}|_{phy}=1968$ MeV. Thus, with Eqs.~(\ref{a0})-(\ref{MKMeta}),
the scattering length $a_0$ can be expressed as a function of
$M_{\pi}$. The four unknown low energy constants $h_2$, $h_3$, $h_4$
and $h_5$ can be determined by fitting the lattice data of
scattering lengths.

Although $(S,I)=(1,1)$ is in fact a couple channel case, we use
single channel unitarization in our fit procedure for all of the
four channels, since so far lattice data only exist without channel
coupling. The obtained LECs in three cases are shown in
Tab.~\ref{tab:LECs}, corresponding to different values of $a(m_D)$
which were estimated by comparing the dispersion relation method
with the cut-off method~\cite{Guo:06}. $a(m_D)$ chosen to be
$-0.373$, $-0.630$ and $-0.864$ correspond to the resultant cut-off
momentum $q_{max}$ at 0.6, 0.8 and 1.0 GeV, respectively. The
obtained scattering length at physical pion mass for each channel
are listed in Tab.~III. In Fig.~1, we plot the fitted results of the
scattering lengths versus pion mass with $a(m_D)=-0.864$ since all
the three choices of $a(m_D)$ give similar curves. One can see that
lattice data can be fitted very well except the $(S,I)=(2,1/2)$
channel. The scattering lengths of $(S,I)=(0,3/2)$ and $(S,I)=(1,1)$
channels are sensitive to pion mass which can be understood from
Tab. 1. In the chiral limit, both of them go to zero. The scattering
lengths of $(S,I)=(-1,1)$ and $(S,I)=(2,1/2)$ channels change little
with the increasing pion mass. The calculation with the heavy meson
chiral perturbation theory is not completely consistent with the
lattice data of $(S,I)=(2,1/2)$ channel which deserves further
investigation. Without the large $N_C$ suppression terms, the
scattering length of $(S,I)=(1,1)$ channel remains zero with the
increasing pion mass. In Ref.~\cite{Guo:09}, with their notation,
the coefficient $C_1=2M_{\pi}^2$ makes the scattering lengths of
$(S,I)=(1,1)$ channel be always negative and decrease with
increasing pion mass. The inclusion of $N_C$ suppression terms
improves the fit of lattice data. Due to the inaccuracy of lattice
data, we show the uncertainty of the scattering lengths in Fig.~1.
The corresponding error bars of the low energy constants are also
listed in Tab.~II. For $h_2$ and $h_4$, the errors are about $10\%-20\%$
and their signs keep the same with all the three parameter sets.
While for $h_3$ and $h_5$, their signs could change with different
parameter sets and the error bars are larger.

\begin{figure}[t]
\begin{minipage}[h]{0.45\textwidth}\centering \subfigure[ ]
{\label {b}\includegraphics[scale=0.3]{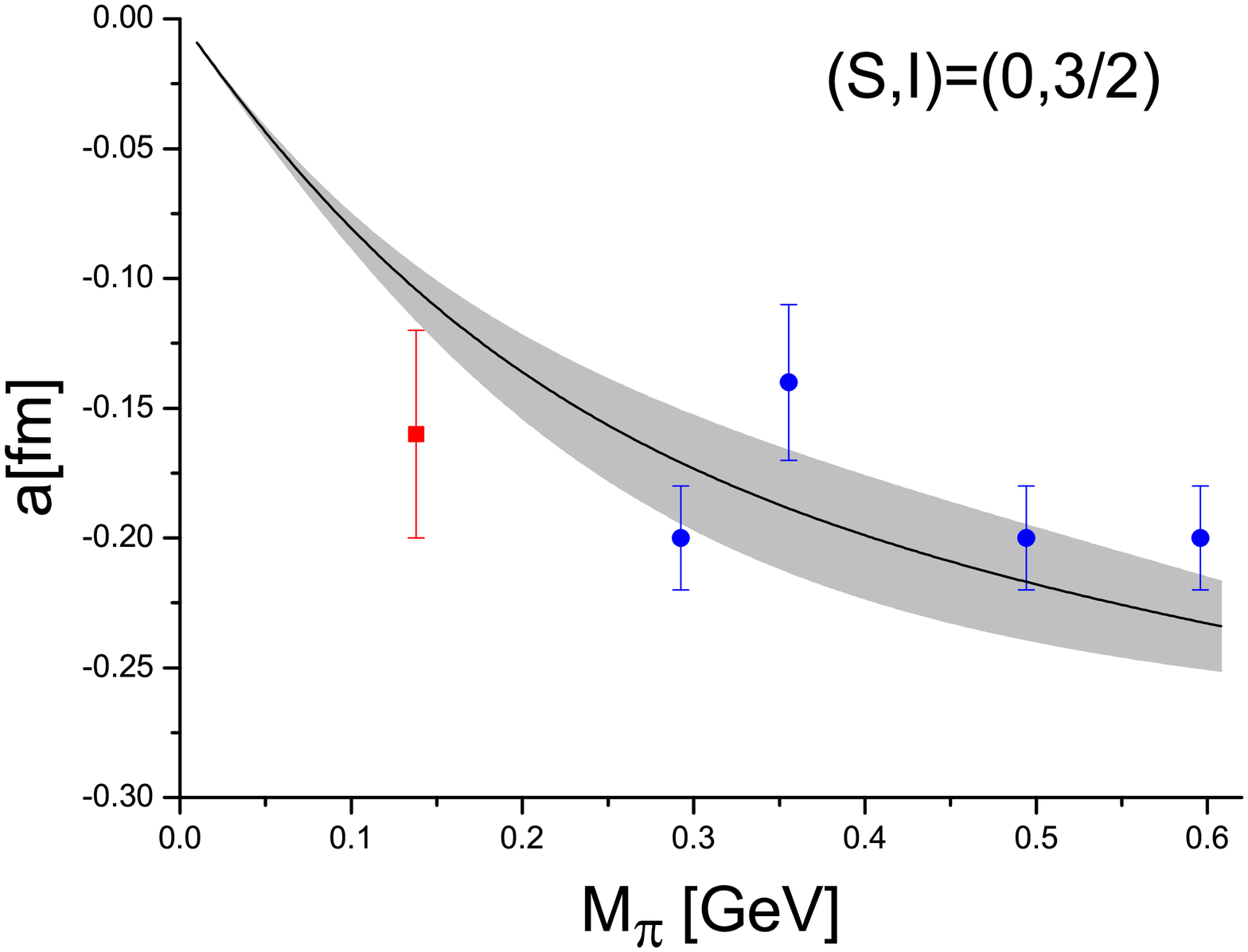}} \end{minipage}
\begin{minipage}[h]{0.45\textwidth}\centering \subfigure[ ]
{\label {b}\includegraphics[scale=0.3]{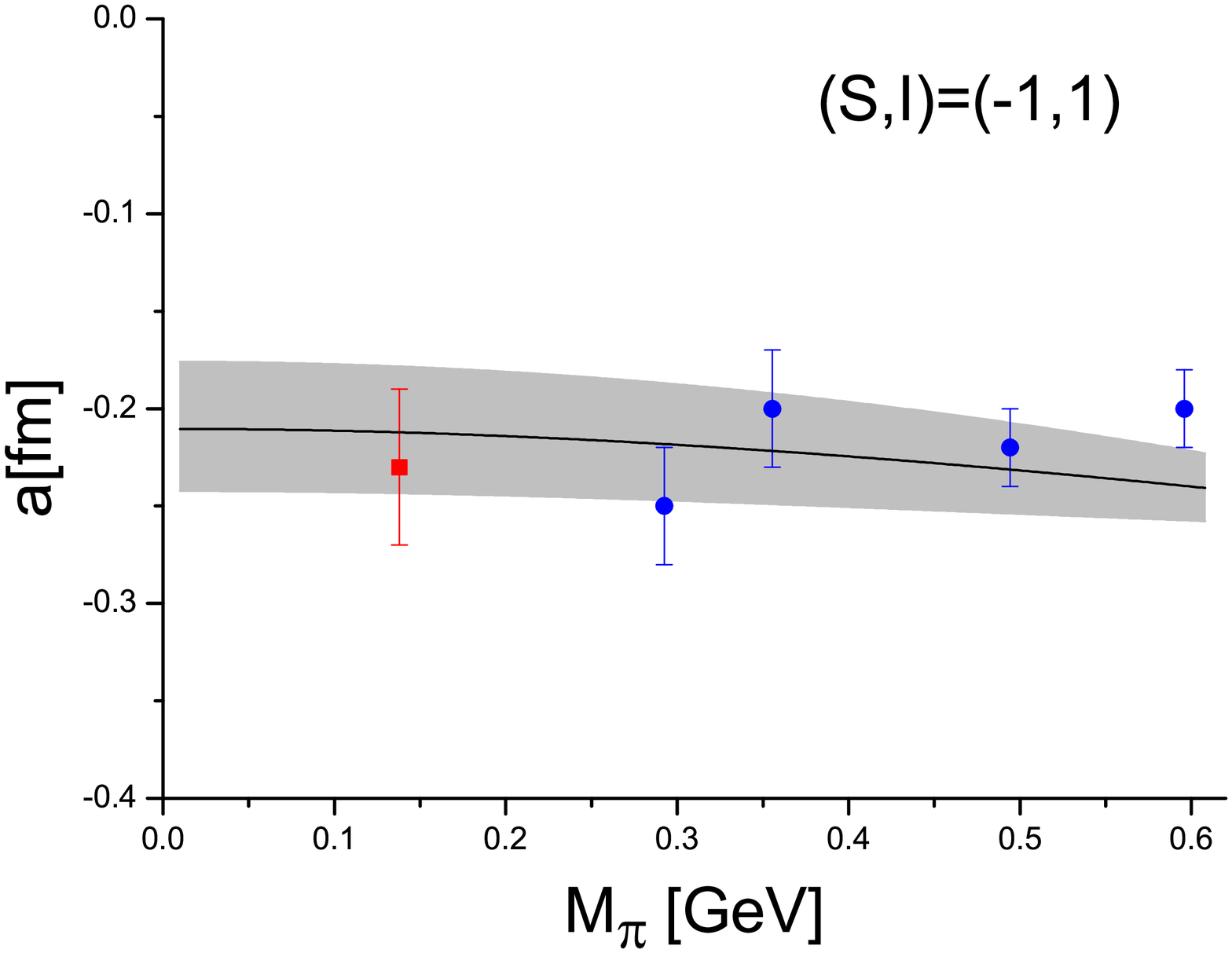}} \end{minipage}
\begin{minipage}[h]{0.45\textwidth}\centering \subfigure[ ]
{\label {b}\includegraphics[scale=0.3]{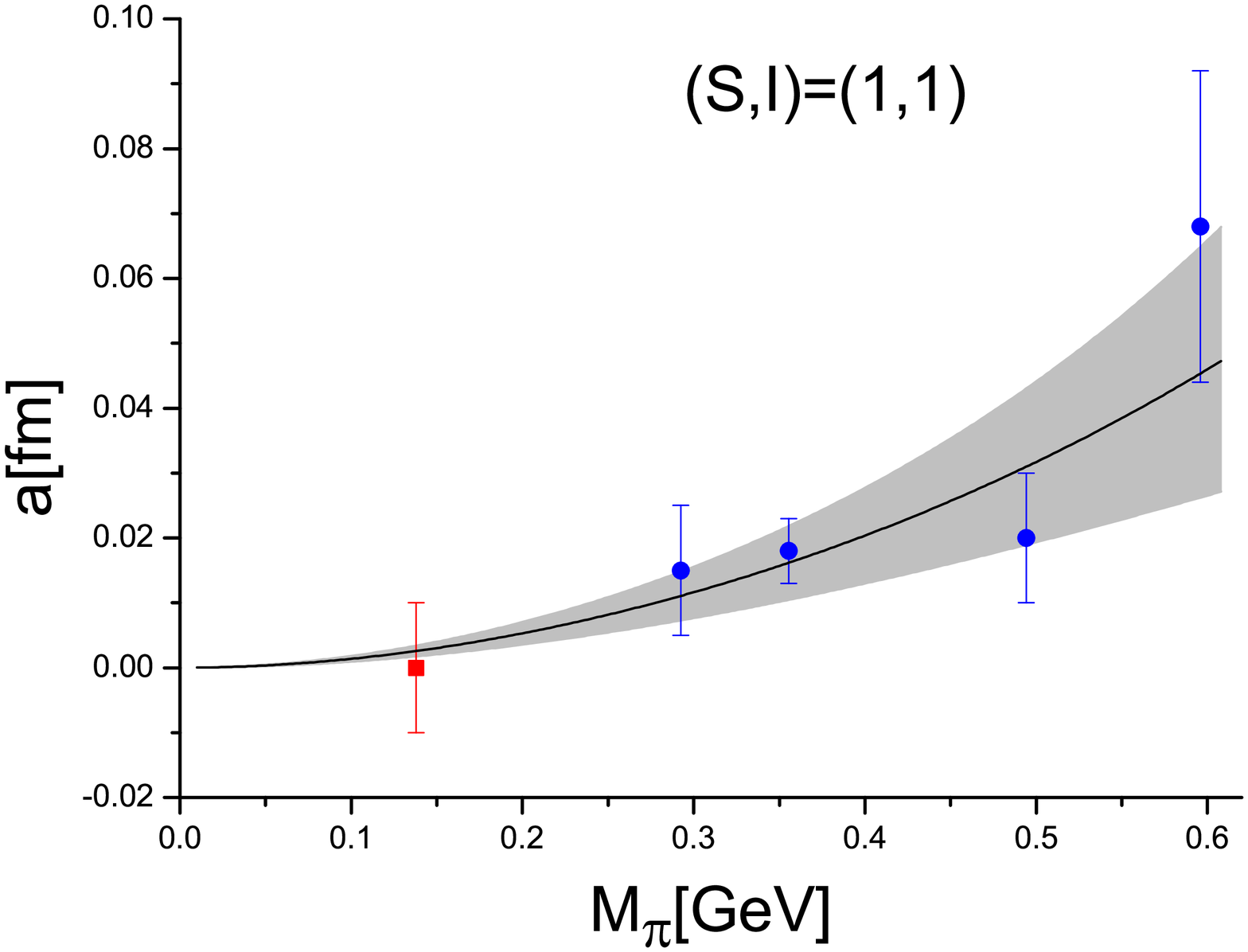}} \end{minipage}
\begin{minipage}[h]{0.45\textwidth}\centering \subfigure[ ]
{\label {b}\includegraphics[scale=0.3]{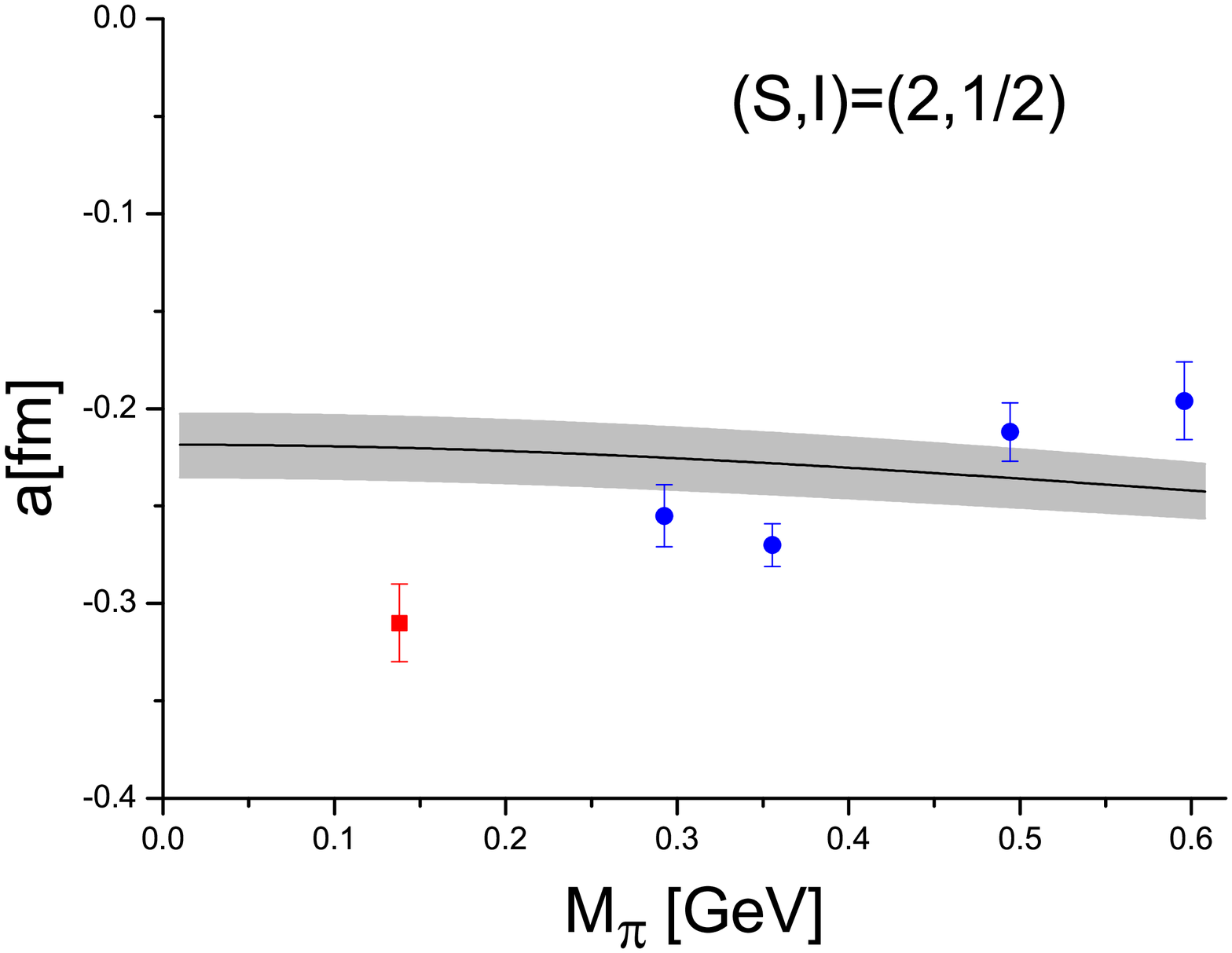}} \end{minipage}
\caption{The fitted results on scattering lengths. The squares at physical pion mass
denote chiral extrapolation results given by~\cite{Liu:2008rz},
which are {\em not} included in our fit. The LECs are taken from fit
III. Solid lines correspond to the central value of each parameter.}
\end{figure}

\section{Pole analysis}
Under the present convention, the relationship between $S$ matrix
and $T$ matrix is given by
\begin{equation}
S_{ij}=\delta_{ij}-\frac{2i}{8\pi}\frac{\sqrt{k_i k_j}}{\sqrt{s}}T_{ij}(s)\ ,
\end{equation}
with $k_i$ the $i$-th channel momentum. In general, for a system
with $N$ open channels, we have in total $2^N$ Riemann sheets, which
can be enumerated as $L(\sigma_1,\sigma_2,\ldots,\sigma_N)$ where
$\sigma_i$ stands for the sign of
$\mathrm{Im}k_i$~\cite{Badalyan:82}. Taking 2-channel case for
example, we enumerate the 4 sheets in the following way,
\begin{eqnarray}
\text{sheet I}: & \ & \mathrm{Im}k_1>0,\ \mathrm{Im}k_2>0,\ L(++)\ ,\non\\
\text{sheet II}: & \ & \mathrm{Im}k_1<0,\ \mathrm{Im}k_2>0,\ L(-+)\ ,\non\\
\text{sheet III}: & \ & \mathrm{Im}k_1<0,\ \mathrm{Im}k_2<0,\ L(--)\ ,\non\\
\text{sheet IV}: & \ & \mathrm{Im}k_1>0,\ \mathrm{Im}k_2<0,\ L(+-)\ .
\end{eqnarray}
The analytic continuation of $S$ matrix to different sheets can be obtained
\begin{equation}
S^{II}=\left(
         \begin{array}{cc}
           \frac{1}{S_{11}} & \frac{iS_{12}}{S_{11}} \\
           \                &   \  \\
           \frac{iS_{12}}{S_{11}} & \frac{\det S}{S_{11}} \\
         \end{array}
       \right),\ \
S^{III}=\left(
         \begin{array}{cc}
           \frac{S_{22}}{\det S} & \frac{-S_{12}}{\det S} \\
           \                     &  \                     \\
           \frac{-S_{12}}{\det S} & \frac{S_{11}}{\det S} \\
         \end{array}
       \right),\ \
S^{IV}=\left(
         \begin{array}{cc}
           \frac{\det S}{S_{22}} & \frac{-iS_{12}}{S_{22}} \\
             \                   &  \                      \\
           \frac{-iS_{12}}{S_{22}} & \frac{1}{S_{22}} \\
         \end{array}
       \right),\ \
\end{equation}
from which we can see that the poles on sheet-II and sheet-III
correspond to zeroes of $S_{11}(s)$ and
$\det{S}=S_{11}S_{22}-S_{12}S_{21}$, respectively.

\begin{table}
\begin{center}
\renewcommand{\arraystretch}{1.5}{\scriptsize
 \begin{tabular}  {|ccc|c|lll|}\hline
 (S,I)                &        Channel                &    Thr   &  RS  &    Fit I       &       Fit II      &   Fit III       \\ \hline
 (-1,0)               &  $D{\bar K}\to D{\bar K}$     &    2363  &   I  &                &         \         &   $2340^{+20}_{-24}$          \\
   \                  &          \                    &    \     &  II  &   $2315^{+30}_{-83}-i70^{+78}_{-70}$   &      $2194^{+117}_{-50}$       &       \         \\
 (0,$\frac{1}{2}$)    &  $D\pi\to D\pi$               &    2005  &  II  &   $2146^{+9}_{-8}-i124^{+14}_{-10}$  &     $2122^{+11}_{-9}-i93^{+13}_{-10}$    &   $2104^{+15}_{-13}-i75^{+19}_{-13}$    \\
   \                  &  $D\eta\to D\eta$             &    2415  & III  &   $2478^{+28}_{-14}-i23^{+5}_{-5}$   &     $2434^{+9}_{-7}-i19^{+11}_{-7}$    &   $2376^{+15}_{-8}-i2^{+17}_{-1}$     \\
   \                  &  $D_s{\bar K}\to D_s{\bar K}$ &    2464  &  \   &      \         &         \         &       \         \\
 (1,0)                &  $DK\to DK$                   &    2363  &  I   &   $2356^{+6}_{-9}$       &      $2327^{+23}_{-19}$       &   $2295^{+40}_{-38}$        \\
   \                  &  $D_s\eta\to D_s\eta$         &    2516  &  \   &     \          &         \         &       \         \\
 (1,1)                &  $D_s\pi\to D_s\pi$           &    2106  &  II  &   $2433^{+50}_{-31}-i26^{+6}_{-9}$   &     $2372^{+38}_{-25}-i39^{+2}_{-6}$    &   $2318^{+39}_{-28}-i37^{+2}_{-3}$    \\
   \                  &  $DK\to DK$                   &    2363  &  \   &     \          &         \         &       \         \\ \hline
 \end{tabular}}
 \caption{\label{pole}\small Pole positions on $\sqrt{s}$ plane in unit of $\mathrm{MeV}$. Thr and RS denote channel threshold and Riemann Sheet, respectively.}
 \end{center}
\end{table}

Corresponding to each set of parameters given by
Tab.~\ref{tab:LECs}, we list the pole positions found in appropriate
channels in Tab.~\ref{pole}, from which one can see all the three
parameter sets give similar results except for the $(S,I)=(-1,0)$
channel. The error bars of the mass and width come from the
uncertainty of the low energy constants $h_i$.

In $(S,I)=(-1,0)$ channel, the pole structure is unstable. Pole
positions are dependent on the strength of interactions, which is
governed by LECs. In fit I, we find a ``resonance", whose position
is similar to~\cite{Guo:09}. However, it has no physical
correspondence since a particle with mass below the lowest
hadron-hadron threshold can not possess finite width by strong
decay. In fit II, there is a virtual state, which is located on the
real axis below $D\bar{K}$ threshold on the second Riemann sheet.
The bound state pole predicted by fit III is in agreement
with~\cite{Hofmann:04}. Further experiments on this channel will
determine which parameter set is more reasonable.

In $(S,I)=(0,1/2)$ channel, we perform 3-channel unitarization. For
example, for the parameter set II, we find a broad second sheet pole
at $(2122_{-9}^{+11}-i93_{-10}^{+13})\mathrm{MeV}$ and a narrow
third sheet pole at $(2434_{-7}^{+9}-i19_{-7}^{+11})\mathrm{MeV}$,
respectively. Although still deviate from the experimental
data~\cite{Belle:04,FOCUS:04}, our results are in agreement
with~\cite{Hofmann:04,Guo:06}. Ref.~\cite{Guo:07} gave some
arguments to explain why resonances predicted theoretically have not
been observed by experiment. In addition to production rate, finding
a new state experimentally also depends on data measurements and
analysis, which are affected by many factors, such as data
statistics, the background, etc.

A bound state pole of $D_{s0}^{*}(2317)$ in $(S,I)=(1,0)$ channel is
obtained. In Ref.~\cite{Guo:09,Cleven:11}, the bound state of
$D_{s0}^{*}(2317)$ is assumed and its mass is used as an input to
determine the LECs. Here, $D_{s0}^{*}(2317)$ is really obtained from
the analysis of the poles. For example, for the parameter set II,
the mass of $D_{s0}^{*}(2317)$ in our analysis is
$m=2327_{-19}^{+23}\mathrm{MeV}$, which is in agreement
with~\cite{Guo:06}.

In $(S,I)=(1,1)$ channel, only $N_C$ suppressed terms contribute to
the elastic scattering amplitudes, as can be seen from
Tab.~\ref{tab:Vstu}. The nearest resonance pole to the physical
region is located on sheet II at
$(2372_{-25}^{+38}-i39_{-6}^{+2})\mathrm{MeV}$. The position of this
state is different from that obtained in Ref.~\cite{Guo:06} where a
resonance on sheet III with smaller mass and larger width exists.

\begin{figure}[t]%
\begin{minipage}[h]{0.45\textwidth}
\centering \subfigure[ ]{ \label{a}
\includegraphics[width=1.0\textwidth]{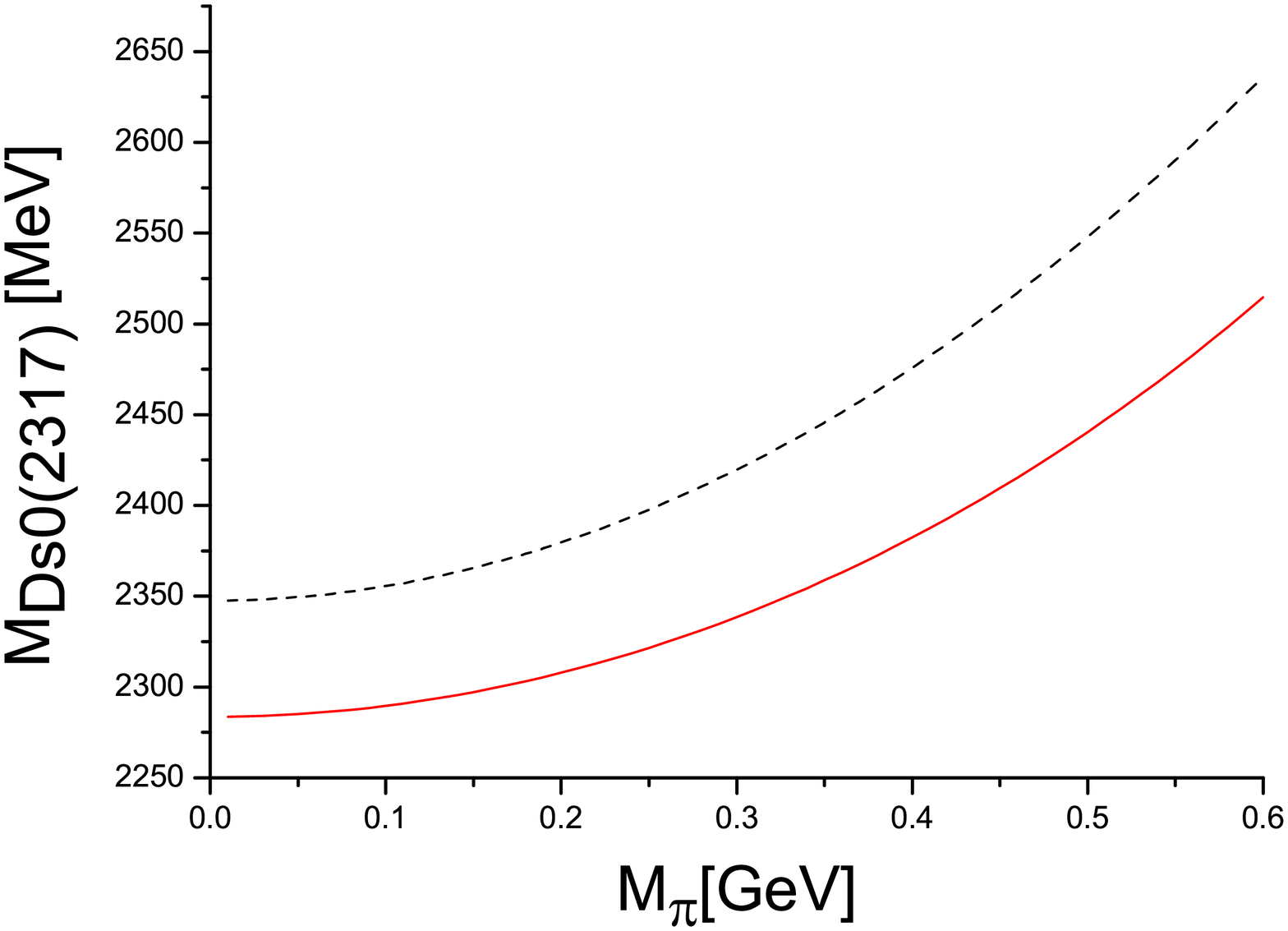}}%
\end{minipage}
\begin{minipage}[h]{0.45\textwidth}
\centering \subfigure[ ]{ \label{b}
\includegraphics[width=1.0\textwidth]{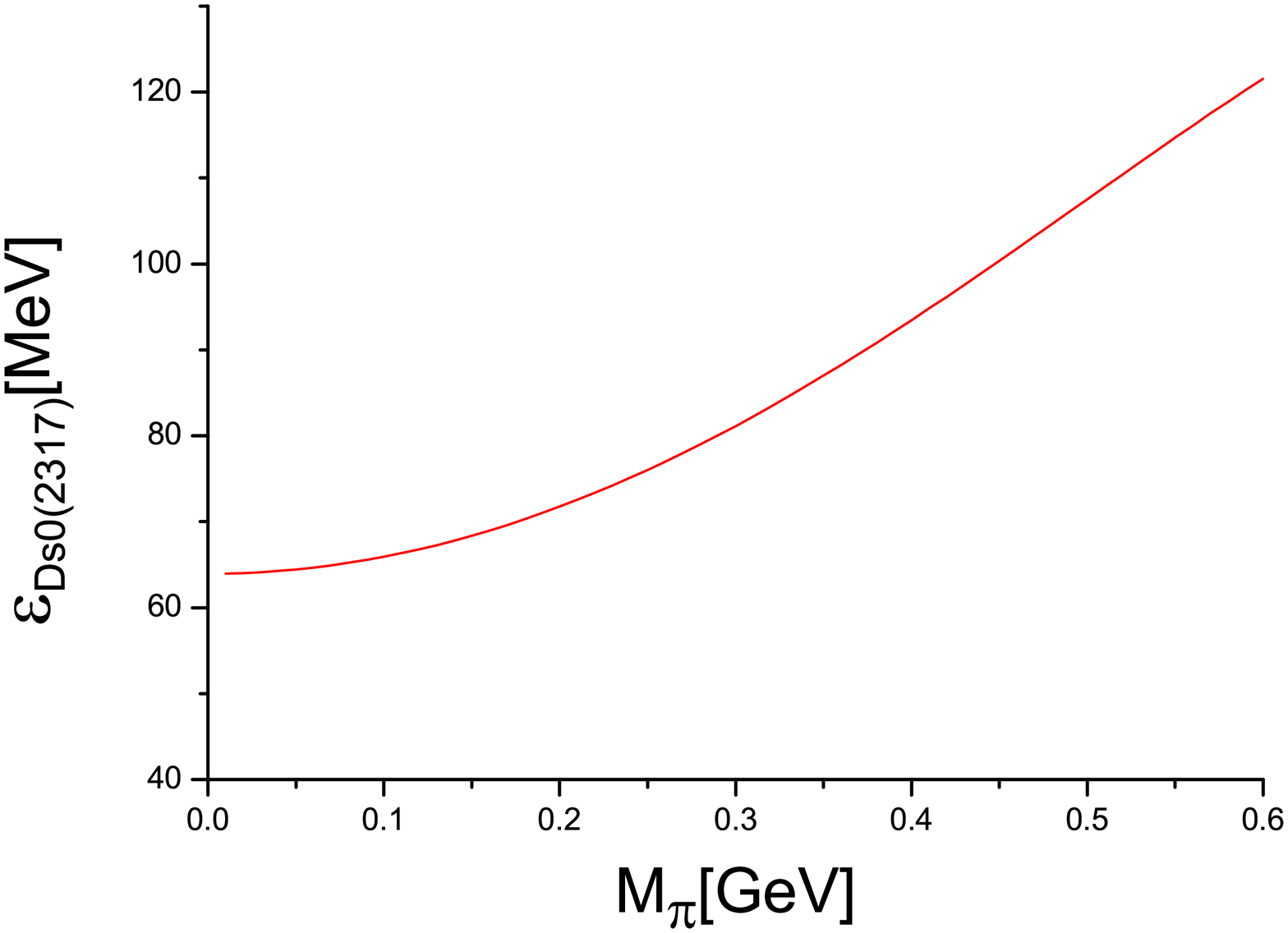}}%
\end{minipage}
\begin{minipage}[h]{0.45\textwidth}
\centering \subfigure[ ]{ \label{c}
\includegraphics[width=1.0\textwidth]{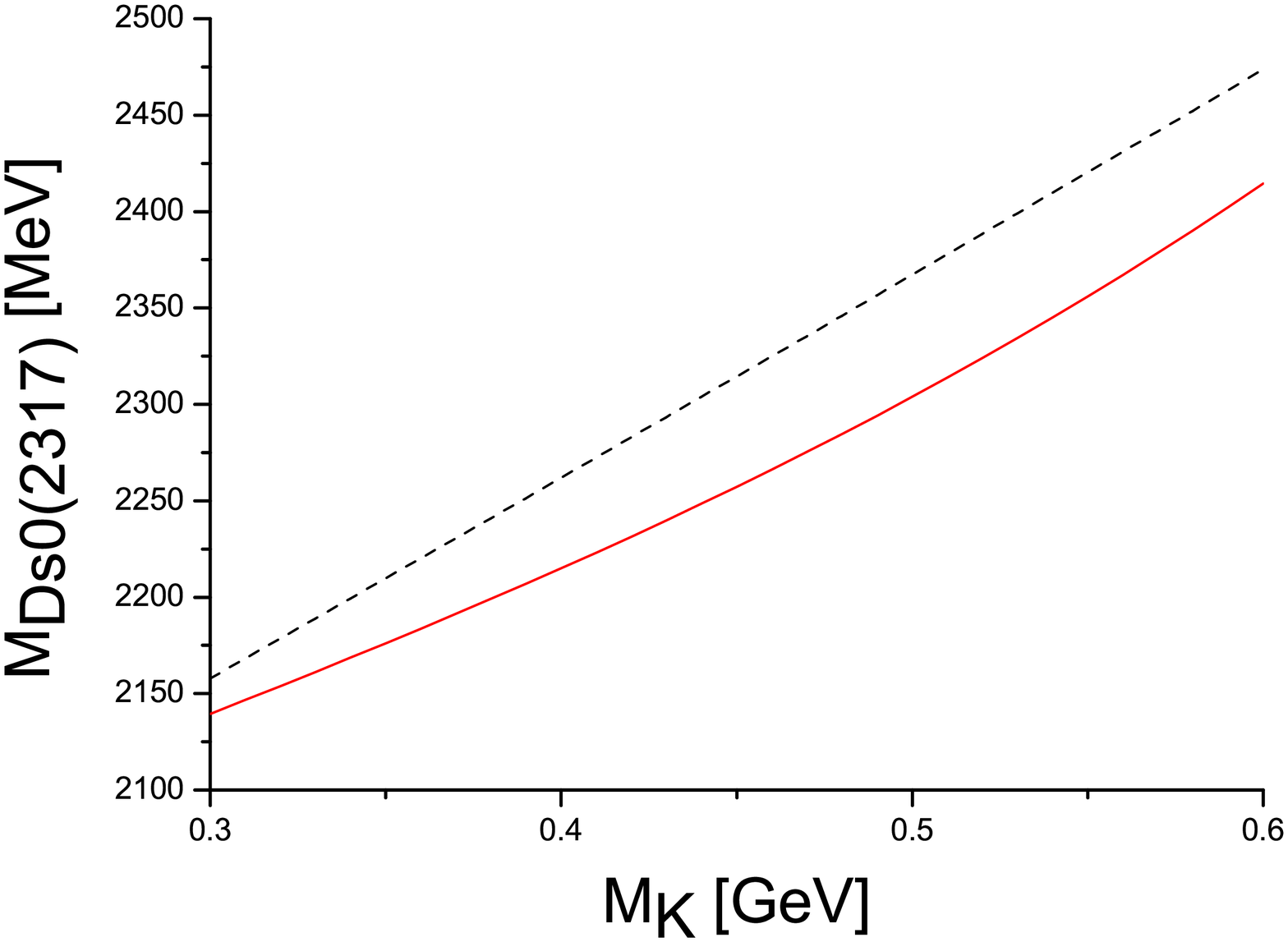}}%
\end{minipage}
\begin{minipage}[h]{0.45\textwidth}
\centering \subfigure[ ]{ \label{d}
\includegraphics[width=1.0\textwidth]{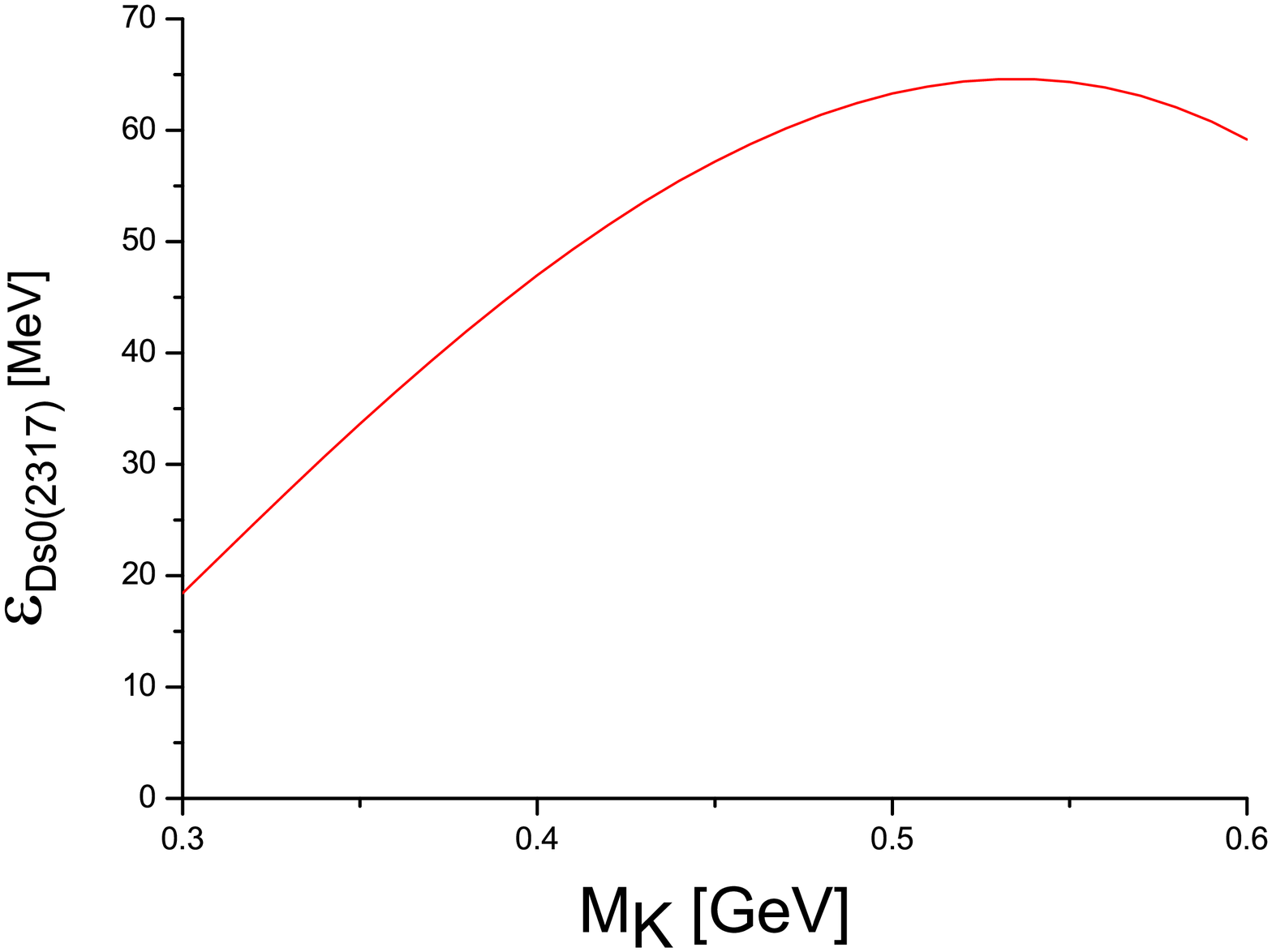}}%
\end{minipage}
\caption{\label{Ds0} {\small a) Pion mass dependence of the mass of
$D_{s0}^{*}(2317)$(solid) and $DK$ threshold(dashed); b) Pion mass
dependence of binding energy; c) Kaon mass dependence of the mass of
$D_{s0}^{*}(2317)$(solid) and $DK$ threshold(dashed); d) Kaon mass
dependence of binding energy.}}
\end{figure}

Although the mass and width of the states are obtained, we did not get the information about the nature of
the states. There are some methods to determine the structure of a
particle, as mentioned in the introduction. In
Ref.~\cite{Cleven:11}, the authors studied the structure by
investigating the quark mass dependence of the states. This method
provides a direct and clear picture for the composition of a
particle. We now make the similar analysis as in Ref.~\cite{Cleven:11}.
We first fix $s$ quark mass and vary the light quark masses. In
Fig.~2a and 2b, we show the mass of $D_{s0}^{*}(2317)$, as well as
the binding energy as a function of the pion mass. Both the mass and
binding energy increase with the increasing pion mass. A pure
$c\bar{s}$ state has no constituent light quarks. Its light quark
mass dependence only comes from sea quark contributions, which
should be very weak as the case of $Ds(1968)$ shown in lattice
simulations~\cite{HPQCD:2008}. The sensitive dependence of light
quark mass shows that $D_{s0}^{*}(2317)$ may probably be a $DK$
molecular or tetraquark state where the constituent light quark
exists.

Some general discussions on the importance of kaon mass dependence
have been made in Ref.~\cite{Cleven:11}. The mass of kaon-hadron
molecular state is given by
\begin{equation}
M=M_K+M_h-\epsilon\ ,
\end{equation}
where $M_h$ is the mass of the other hadron and $\epsilon$ is the
binding energy. The leading kaon mass dependence of such a bound
state is linear, and the slop is unity.

To study the kaon mass dependence of $D_{s0}^{*}(2317)$, we fix the
pion mass at its physical value, and express all results in terms of
$M_K$. From Eq.~(\ref{mass}) and (\ref{mass-relation}), we can
obtain
\begin{eqnarray}
M_D(M_K)&=&M_D|_{phy}+\frac{2h_0}{M_D|_{phy}}(M_{K}^2-M_{K}^2|_{phy})\ ,\non\\
M_{Ds}(M_K)&=&M_{Ds}|_{phy}+\frac{2(h_0+h_1)}{M_{Ds}|_{phy}}(M_{K}^2-M_{K}^2|_{phy})\ ,
\end{eqnarray}
and
$M_{\eta}(M_K)=\sqrt{\frac{4}{3}M_K^2-\frac{1}{3}M_{\pi}^2|_{phy}}$.

Fig.~2c and 2d show the mass and binding energy of
$D_{s0}^{*}(2317)$ as a function of kaon mass. The $K$-meson mass
dependence of $D_{s0}^{*}(2317)$ is almost linear which is really in
good agreement with the $DK$ molecular expectation. These results
are comparable with Refs.~\cite{Cleven:11,Faessler}.

\section{Summary}\label{Summary}

In this paper, we calculate the complete scattering amplitudes of
Goldstone bosons off the pseudoscalar D-mesons using unitarized
heavy meson chiral approach. Two low energy constants $h_0$ and
$h_1$ are determined by the mass splitting among $D$ mesons and pion
mass dependence of $D$ and $D_s$. The other four LECs are determined
by fitting lattice simulations on $S$-wave scattering lengths. The
large $N_C$ suppressed terms improve the fit. Three sets of
parameters are obtained according to different choices of the
subtraction constant $a(m_D)$. All the three sets of parameters give
similar scattering lengths which are close to the lattice results.

For three parameter sets, the positions of the poles in each channel
are close except in $(S,I)=(-1,0)$ channel. In this channel, the
poles are sensitive to the parameter sets. Further experiments or
lattice simulation can determine which parameter set is more
reasonable. For other channels, we take the average values of the
mass and width as the final results. $D_{s0}^{*}(2317)$ is obtained
as a bound state in $(S,I)=(1,0)$ channel, with the mass being
$m=2326_{-22}^{+23}\mathrm{MeV}$. The strong pion mass dependence of
its mass and binding energy disfavors conventional $c\bar{s}$
content. The approximately linear kaon mass dependence reveals it is
predominately a $DK$ molecular state. In $(S,I)=(0,1/2)$ channel, a
broad pole structure is found at
$(2124_{-10}^{+12}-i97_{-11}^{+15})\mathrm{MeV}$ on the second
Riemann sheet, and a narrow pole is at
$(2429_{-10}^{+17}-i15_{-4}^{+11})\mathrm{MeV}$ on the third Riemann
sheet. A resonance pole also exists on the second Riemann sheet with
mass and half width $2374_{-28}^{+42}$ MeV and $34_{-6}^{+3}$ MeV in
$(S,I)=(1,1)$ channel.

\section*{Acknowledgement} We would like to thank L.M.~Liu, H.W.~Lin
and H.Q.~Zheng for helpful communications. This work is supported in part by DFG and NSFC (CRC 110)
and by National Natural Science Foundation of China (Grant No. 11035006).

\end{document}